\documentclass[conference]{IEEEtran}
\IEEEoverridecommandlockouts
\usepackage{cite}
\usepackage{amsmath,amssymb,amsfonts,amsthm}
\newtheorem{theorem}{Theorem}

\usepackage{algorithmic}
\usepackage{graphicx}
\usepackage{textcomp}
\usepackage{makecell}
\usepackage{xcolor}
\usepackage{url}
\def\BibTeX{{\rm B\kern-.05em{\sc i\kern-.025em b}\kern-.08em
    T\kern-.1667em\lower.7ex\hbox{E}\kern-.125emX}}
\begin{document}

\title{Joint Analog Encoder Design for Multi-Task Oriented Wireless Communication
}

\author{\IEEEauthorblockN{Chenmin Sha}
\IEEEauthorblockA{\textit{Department of Electronic Engineering}\\
\textit{Tsinghua University} \\
Beijing, China \\
scm21@mails.tsinghua.edu.cn}
\and
\IEEEauthorblockN{Shidong Zhou$^\dagger$}
\IEEEauthorblockA{\textit{Department of Electronic Engineering}\\
\textit{Tsinghua University} \\
Beijing, China \\
zhousd@mail.tsinghua.edu.cn}
}

\maketitle

\begin{abstract}
In this paper we study multi-task oriented communication system via studying analog encoding method for multiple estimation tasks. The basic idea is to utilize the correlation among interested information required by different tasks and the feature of broadcast channel. For linear estimation tasks, we provide a low complexity design for multi-user multi-task system based on orthogonal decomposition of subspaces. It is proved to be optimal in some special cases, and for general cases, numerical results also show it can achieve near-optimal performance. Further, we make a trial to migrate above method to neural networks based non-linear estimation tasks, and it also shows improvement in energy efficiency.  
\end{abstract}

\begin{IEEEkeywords}
task-oriented, multi-user, multi-task
\end{IEEEkeywords}

\section{Introduction}
Task-oriented communication systems have gained significant attention in recent years, with a focus on ensuring quality of tasks, such as estimation error\cite{arafa2021sample}, classification accuracy\cite{shao2023task}, and control performance, rather than traditional performance metrics like data rate and transmission reliability in wireless sensing or control system. This shift in focus has led to the development of task-oriented or semantic-based scheduling and data compression schemes. 

In addition to optimizing the system for a single task, in a multi-user network, different users may need to perform different tasks, so it is necessary to consider the multi-task oriented communication system to satisfy heterogeneous requirements. Several recent works have paid attention to such case, e.g., \cite{xie2022task,hu2022one}. \cite{xie2022task} and \cite{zhang2022unified} each gives a unified semantic communication framework for handling multiple tasks in one-to-one communication scenario. And \cite{hu2022one} further studies the system design under broadcast case with one transmitter and multiple receivers. But all these works only consider different tasks are performed based on different independent data sources and they aim to coordinate the competition between different tasks for communication resources. 

However, multiple users may demand for different customized services from the same data source. And these tasks, with high probability, are not independent. For example, in a multi-user augmented reality (AR) system, different users may subscribe different views from a global map and their interested contents can be correlated \cite{zhang2022sear}, e.g., different angles of view of the same object.  Unfortunately, existing works mentioned above have overlooked these inherent relationships between tasks. 

Some works \cite{sheng2022multi,zhang2022deep} have explored the scenarios where the same data is required for different tasks but they only consider the single user case, thus neglecting more challenging one-to-many case that involves the varying channel conditions among users. 

\cite{kang2023personalized} has incorporated users' personalized saliency into task-oriented image transmission, but their approach is limited to time division multiplexing and overlooks the properties of broadcast channels as well as the potential overlapping of personal interests. 

Therefore, how to utilize the correlation among the interested information required by different tasks in multi-user communication system design remains unclear.

To address the above problems, we take an initial step by examining how to optimize the analog encoder in analog communication system to serve different estimation tasks for multiple users. Although digital communication is more common these days, the concise in mathematics of analog system can help to derive general conclusions that can guide the design of more complex multi-task oriented communication systems.

Additionally, analog transmission itself has attracted lots of research owing to its low complexity and automatic adaption to channel conditions, and such scheme is widely used in wireless sensor networks, e.g., \cite{leong2011power,ren2017infinite,SurezCasal2017AnalogTO}. These works are not task-oriented or just optimize for single sensing or control task, so they are orthogonal to our work in this way. 



Concentrating on analog encoding for multiple users with customized tasks, our contributions are summarized as follows: 
\begin{itemize}
    \item For multiple linear Gaussian estimation tasks, we propose an encoder design based on subspace decomposition, which is proved to be optimal under certain conditions. Besides, it achieves near-optimal performance with significantly lower complexity compared to the optimal solution provided by us.
    \item In addition, our method has a clear geometric interpretation. We demonstrate that we can represent users' interests using subspaces and extract independent content for all tasks through orthogonal decomposition of these subspaces. Furthermore, we allocate energy to each content based on both its importance and the channel conditions. 
    \item Then we extend to non-linear estimation tasks based on neural networks. We leverage the insights gained from linear case to design the transceiver, which also outperforms baseline methods. 
\end{itemize}

The organization of the paper is as follows: We first introduce system model in Section II, and we especially introduce the Gaussian linear model, which we will study first. In Section III, we obtain the optimal solution to Gaussian case by semi-definite programming. Then in section IV, we provide a low complexity algorithm for solving the problem from the view of subspace. In section V, we turn to non-linear estimation tasks and show how the conclusions obtained in linear case can be applied to the non-linear problem. The numerical results are presented in Section V and we conclude our work and point out future directions in Section VI. 

\section{System Model}
\subsection{General System Architecture}
We are considering a system that includes a single data source (i.e., sensor) with high dimensional source data $\mathbf{y}\in\mathbb{R}^{D_y}$, serving $N$ users through radio channel. We suppose the task of user $n$ is to obtain information about a certain vector $\mathbf{x}_n\in\mathbb{R}^{D_x}$ that is related to $\mathbf{y}$. Due to the communication noise and limited communication resources, the sensor needs to encode $\mathbf{y}$ and broadcast
\begin{equation}
    \mathbf{s}=\mathbf{g}(\mathbf{y})
\end{equation}
and $\mathbf{g}(\cdot)$ is the encoder at sensor side. Then user $n$ will receive
\begin{equation}
    \mathbf{r}_n=h_n\mathbf{s}+\mathbf{u}_n
\end{equation}
where $h_n\in\mathbb{R}$ is the block fading channel between sensor and user $n$ and we only consider real scalar channel here for simplicity. And $\mathbf{u}_n$ is the additive noise at receiver side and we assume $\mathbf{u}_n\sim\mathcal{N}(\mathbf{0},\mathbf{I})$. Besides, we assume the channel is known by both sides. 

After receiving the signal, the user will extract relevant information using a decoder $\mathbf{f}_n(\cdot)$ to get an estimated value
\begin{equation}
    \hat{\mathbf{x}}_n=\mathbf{f}_n(\mathbf{r}_n)
\end{equation}

\begin{figure*}[ht]
    \centering
    \includegraphics[width=0.8\linewidth]{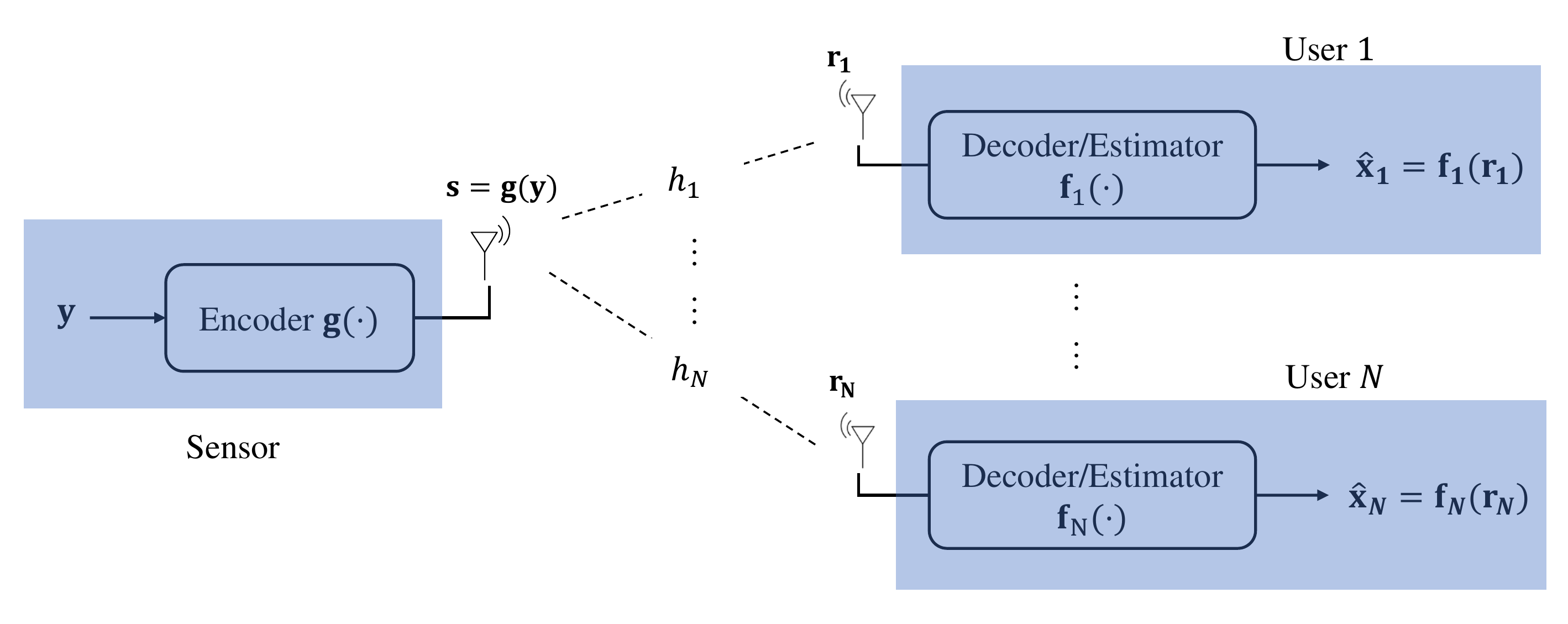}
    \caption{System Model for Multiple Estimation Tasks}
    \label{systemmodel}
\end{figure*}

The general system architecture is shown in Fig. \ref{systemmodel}. We choose the mean squared error (MSE) between the reconstructed value $\hat{\mathbf{x}}_n$ and the ground truth $\mathbf{x}_n$ as the evaluation metric for the tasks. And our goal is to design the encoder and decoders to minimize the sum of all users' MSE under the transmission energy constraint, i.e.,
\begin{equation}\label{eq:general_prob}
    \begin{aligned}
        \min_{\mathbf{g}(\cdot),\{\mathbf{f}_n(\cdot)\}} \quad & \sum_{n=1}^N\operatorname{E}[\Vert\mathbf{x}_n-\hat{\mathbf{x}}_n\Vert^2_2] \\
        \text{s.t.} \quad & \operatorname{E}[\Vert\mathbf{s}\Vert_2^2]\leq E
    \end{aligned}
\end{equation}

\subsection{Gaussian Linear Model}
A special case is when the variables are jointly Gaussian, and the encoder and decoders are all linear operations. We first pay attention to such case to gain useful intuition. Assume that $\mathbf{x}_n$ and $\mathbf{y}$ follows joint Gaussian distribution, i.e.,
\begin{equation}
\begin{aligned}
    & \begin{pmatrix}
	\mathbf{x}_n \\
	\mathbf{y}
    \end{pmatrix}\sim\mathcal{N}(\begin{pmatrix}
    \boldsymbol{\mu}_{x_n} \\
    \boldsymbol{\mu}_y
    \end{pmatrix},\begin{pmatrix}
    \boldsymbol{\Sigma}_{x_n} & \boldsymbol{\Sigma}_{x_ny}\\
    \boldsymbol{\Sigma}_{yx_n} & \boldsymbol{\Sigma}_y
    \end{pmatrix}), \\
    & \quad n = 1,\cdots,N
\end{aligned}
\end{equation}
and without loss of generality, we can assume $\boldsymbol{\Sigma}_y=\mathbf{I}$. In linear case, the encoder and decoders are represented by $\mathbf{G}$ and $\{\mathbf{F}_n\}$ respectively. That is, user will receive 
\begin{equation}
    \mathbf{r}_n=h_n\mathbf{s}+\mathbf{u}_n=h_n\mathbf{Gy}+\mathbf{u}_n
\end{equation}
and obtains estimation by
\begin{equation}
    \hat{\mathbf{x}}_n=\mathbf{F}_n\mathbf{r}_n
\end{equation}

Under such setting, the optimal linear estimator given $\mathbf{G}$ is:
\begin{equation}\label{optimal_estimator}
    \begin{aligned}
        \mathbf{F}_n^* = h_n\boldsymbol{\Sigma}_{x_ny}\mathbf{G}^\top(h_n^2\mathbf{G}\mathbf{G}^\top+\mathbf{I})^{-1}
    \end{aligned}
\end{equation}
with the MSE as
\begin{equation}\label{MSE}
\begin{aligned}
     & \operatorname{Tr}(\boldsymbol{\Sigma}_{x_n}-\boldsymbol{\Sigma}_{x_ny}\mathbf{G}^\top(\mathbf{G}\mathbf{G}^\top+h_n^{-2}\mathbf{I})^{-1}\mathbf{G}\boldsymbol{\Sigma}_{yx_n}) \\
     = & \text{MSE}_n^*+\operatorname{Tr}(h_n^{-2}\boldsymbol{\Sigma}_{yx_n}\boldsymbol{\Sigma}_{x_ny}(\mathbf{G}^\top\mathbf{G}+h_n^{-2}\mathbf{I})^{-1})  
\end{aligned}
\end{equation}
where $\text{MSE}_n^*=\operatorname{Tr}(\boldsymbol{\Sigma}_{x_n}-\boldsymbol{\Sigma}_{x_ny}\boldsymbol{\Sigma}_{yx_n})$ is the MSE without communication noise. And the latter term represents additional error in $\mathbf{x}_n$ caused by estimation error of $\mathbf{y}$ at user side. 

And in Gaussian linear setting, the original optimization problem \eqref{eq:general_prob} has following form:
\begin{equation}\label{P2}
    \begin{aligned}
        \min_{\mathbf{G}}  & \quad \sum_{n=1}^N \operatorname{Tr}(h_n^{-2}\mathbf{M}_n(\mathbf{G}^\top\mathbf{G}+h_n^{-2}\mathbf{I})^{-1}) \\
        \text{s.t.}&\quad \operatorname{Tr}(\mathbf{G}^\top\mathbf{G})\leq E
    \end{aligned}
\end{equation}
where $\mathbf{M}_n=\boldsymbol{\Sigma}_{yx_n}\boldsymbol{\Sigma}_{x_ny}$.

\section{Optimal Encoder for Gaussian Linear Case}
We first provide a solution \eqref{P2} that is solved through semi-definite programming (SDP), and is ensured to be optimal. By introducing auxiliary variables, we can turn the problem into
\begin{equation}
	\begin{aligned}
		\min \quad & \sum_{n=1}^N\operatorname{Tr}(\mathbf{M}_n\mathbf{Q}_n) \\
		\text{s.t.} \quad & \operatorname{Tr}(\mathbf{R})\leq E \\
		& \mathbf{Q}_n\succeq (h_n^2\mathbf{R}+\mathbf{I})^{-1}, \ n=1,\cdots,N \\
        & \mathbf{R}\in\mathbb{S}^{D_y}_+, \ \mathbf{Q}_n\in\mathbb{S}^{D_y}_+, \ n=1,\cdots,N
	\end{aligned}
\end{equation}
Then using Schur complete, we can transform the above problem into the following SDP problem:
\begin{equation}
	\begin{aligned}
		\min \quad & \sum_{n=1}^N\operatorname{Tr}(\mathbf{M}_n\mathbf{Q}_n) \\
		\text{s.t.} \quad & \operatorname{Tr}(\mathbf{R})\leq E \\
		& \begin{pmatrix}
			\mathbf{Q}_n & \mathbf{I} \\
			\mathbf{I} & h_n^2\mathbf{R}+\mathbf{I}
		\end{pmatrix} \succeq \mathbf{0}, \ n=1,\cdots,N\\
        & \mathbf{R}\in\mathbb{S}^{D_y}_+, \ \mathbf{Q}_n\in\mathbb{S}^{D_y}_+, \ n=1,\cdots,N
	\end{aligned}
\end{equation}
which can be solved directly using convex optimization tools off-the-shelf, e.g., CVX \cite{cvx,gb08}. And we can obtain optimal $\mathbf{G}$ by decomposing $\mathbf{R}=\mathbf{G}^\top\mathbf{G}$. 

However, such optimization cannot provide useful insights that can extend to more complex scenarios, which goes against our original purpose of studying the problem. The results obtained from this method will just serve as a reference and in following sections, we will focus on looking for more informative solutions. 

\section{Finding Solutions with Geometric Illustration}
Look back at the problem \eqref{P2}, since it depends on $\mathbf{G}$ only through $\mathbf{G}^\top\mathbf{G}$, we can assume that $\mathbf{G}=\sqrt{\mathbf{W}}\mathbf{P}^\top$ where 
$$
\begin{aligned}
    & \mathbf{W}=\operatorname{diag}(w_1,\cdots,w_{D_y}),\ w_1,\cdots,w_{D_y}\geq 0 \\
    & \mathbf{P}=(\mathbf{p}_1,\cdots,\mathbf{p}_{D_y}),\ \mathbf{p}_i^\top\mathbf{p}_j=\delta_{ij}
\end{aligned}
$$
Such structure can be explained as first projecting data onto some space then allocating energy among different dimensions of projected data --- and we have to determine best projection space and energy allocation scheme. To solve these two problems, we begin by examining the single user case to gain some useful insights on the optimal projection, and then extend it to multi-user multi-task case. 

\subsection{Optimal Solution in Single-User Case}
In single-user case, \eqref{P2} can be written in the following form based on \eqref{MSE}: 
\begin{equation}\label{prob:singleuser}
    \begin{aligned}
        \min_{\mathbf{G}}  \quad & \operatorname{Tr}(\boldsymbol{\Sigma}_{yx}\boldsymbol{\Sigma}_{xy}(\mathbf{G}^\top\mathbf{G}+h^{-2}\mathbf{I})^{-1}) \\
        \text{s.t.} \quad & \operatorname{Tr}(\mathbf{G}^\top\mathbf{G})\leq E
    \end{aligned}
\end{equation}
whose optimal solution to \eqref{prob:singleuser} is given as follows: 
\begin{theorem}
The optimal MSE in single-user case can be achieved by linear encoder:
\begin{equation}
\mathbf{G}=\sqrt{\mathbf{W}}\mathbf{P}^\top=\begin{pmatrix}
    \sqrt{w_1} & & \\
    & \ddots & \\
    & & \sqrt{w_{D_y}}
\end{pmatrix}\begin{pmatrix}
    \mathbf{p}_1^\top \\
    \vdots \\
    \mathbf{p}_{D_y}^\top
\end{pmatrix}
\end{equation}
where $w_d=\max(\sqrt{\lambda_d/\beta}-h^{-2},0),d=1,\cdots,D_y$ and $\mathbf{p}_d$ is eigenvector of $\boldsymbol{\Sigma}_{yx}\boldsymbol{\Sigma}_{xy}$ with corresponding eigenvalue $\lambda_d$. 
\end{theorem}
\begin{IEEEproof}
We refer to the Lagrange multiplier method to find the optimal solution. We introduce a Lagrange multiplier $\beta>0$ to enforce the energy constraint $\text{Tr}(\mathbf{G}^\top\mathbf{G}) \leq E$, and the Lagrangian function is 
\begin{equation}
    \mathcal{L}_{\beta}(\mathbf{G}) = \operatorname{Tr}(\mathbf{M}(\mathbf{G}^\top\mathbf{G}+h^{-2}\mathbf{I})^{-1})+\beta\text{Tr}(\mathbf{G}^\top\mathbf{G})
\end{equation}
Taking the partial derivative of $\mathcal{L}_{\beta}(\mathbf{G})$ with respect to $\mathbf{G}$ and setting it to zero, we get the following expression, which provides necessary condition for optimal solutions. 
\begin{equation}\label{eq:nec}
   \mathbf{G}(\mathbf{G}^\top\mathbf{G}+h^{-2}\mathbf{I})^{-1}\mathbf{M} = \beta\mathbf{G}(\mathbf{G}^\top\mathbf{G}+h^{-2}\mathbf{I})
\end{equation}
Suppose there exists a $\mathbf{G}=\sqrt{\mathbf{W}}\mathbf{P}^\top$ satisfying \eqref{eq:nec}, then we have:
\begin{equation}
\sqrt{\mathbf{W}}(\mathbf{W}+h^{-2}\mathbf{I})^{-1}\mathbf{P}^\top\mathbf{M}=\beta\sqrt{\mathbf{W}}(\mathbf{W}+h^{-2}\mathbf{I})\mathbf{P}^\top
\end{equation}
And this indicates that when $w_d>0$, then $\mathbf{p}_d^\top$ must be an eigenvector of $\mathbf{M}$. So we can choose $\mathbf{P}^\top$ as eigenvectors of $\mathbf{M}$, i.e.,
$$
\mathbf{P}^\top\mathbf{M}=\boldsymbol{\Lambda}\mathbf{P}^\top
$$
with $\boldsymbol{\Lambda}=\operatorname{diag}(\lambda_1,\cdots,\lambda_{D_y})$. When the eigenvalues are all different, $\mathbf{P}$ is unique if ignoring the order. Substituting such $\mathbf{G}=\sqrt{\mathbf{W}}\mathbf{P}^\top$ into \eqref{eq:nec} and we get 
\begin{equation}
     \sqrt{\mathbf{W}}(\mathbf{W}+h^{-2}\mathbf{I})^{-1}\boldsymbol{\Lambda}=\beta\sqrt{\mathbf{W}}(\mathbf{W}+h^{-2}\mathbf{I})
\end{equation}which yields the solution set:
$$
w_d\in\{0,\max(\sqrt{\lambda_d/\beta}-h^{-2},0)\}\quad d=1,\cdots,D_y
$$
Each of them leads to a local optima of $\mathcal{L}_\beta(\mathbf{G})$. And it is obvious to see the global optima is achieved when choosing $w_d$ as 
$$
w_d=\max(\sqrt{\lambda_d/\beta}-h^{-2},0)\quad d=1,\cdots,D_y
$$
Besides, we can verify that different forms of $\mathbf{P}$ won't affect the final function value. By now we have completed the proof. 
\end{IEEEproof}
Note that based on above theorem, the actual dimension of encoder output $\mathbf{s}$ is no more than the rank of $\boldsymbol{\Sigma}_{xy}$, which is an explicit indicator of the benefit of consider specific requirement of the task in system design. Besides, the eigenvectors of $\boldsymbol{\Sigma}_{yx}\boldsymbol{\Sigma}_{xy}$ are just right singular vectors of $\boldsymbol{\Sigma}_{xy}$, which are also a set of orthonormal basis of the row space of $\boldsymbol{\Sigma}_{xy}$.

Recall that the MMSE estimation of $\mathbf{x}$ given perfect reception of $\mathbf{y}$ is 
\begin{equation}
	\hat{\mathbf{x}}_n=\boldsymbol{\Sigma}_{xy}\mathbf{y}
\end{equation}
so $\boldsymbol{\Sigma}_{xy}$ can represent the part of information by the task, and its row space can be seen as the ``interested" subspace. Our optimal projection matrix just computes the projection of data onto such subspace. 



\subsection{Extension to Multiple Users}
Then we turn our sight to the multi-user multi-task case and we begin with searching for the optimal projection space. As single user case \eqref{eq:nec}, we can write the corresponding necessary condition for optimality:
\begin{equation}\label{multiuser_nec}
    \sum_{n=1}^N\mathbf{G}(\mathbf{G}^\top\mathbf{G}+h_n^{-2}\mathbf{I})^{-1}h_n^{-2}\mathbf{M}_n(\mathbf{G}^\top\mathbf{G}+h_n^{-2}\mathbf{I})^{-1}=\beta\mathbf{G}
\end{equation}

In general, it is hard to find solutions to \eqref{multiuser_nec}. However, in special cases, we can provide a solution according to the following theorem.  
\begin{theorem}\label{mutliuser_optimal}
    Denote the set of eigenvectors of $\sum_{n=1}^N\mathbf{M}_n$ corresponding to positive eigenvalues as $\mathcal{S}$, and the set of $\mathbf{M}_n$ as $\mathcal{S}_n$. If $\mathcal{S}_n\subseteq\mathcal{S}, \forall n$, then a solution to \eqref{multiuser_nec} can be found by setting $\mathbf{P}$ as eigenvectors of $\sum_{n=1}^N\mathbf{M}_n$ and $\mathbf{G}=\sqrt{\mathbf{W}}\mathbf{P}^\top$. And thus we found at least a local optima to \eqref{P2}. 
\end{theorem}
\begin{IEEEproof}
When $\sum_{n=1}^N\mathbf{M}_n=\mathbf{P}\boldsymbol{\Lambda}\mathbf{P}^\top$, based on the assumptions, we can write each $\mathbf{M}_n$ in the form $\mathbf{M}_n=\mathbf{P}\boldsymbol{\Lambda}^{(n)}\mathbf{P}^\top$, where $\boldsymbol{\Lambda}$ and all $\boldsymbol{\Lambda}^{(n)}$ are all diagonal matrix, and denote $\boldsymbol{\Lambda}^{(n)}=\operatorname{diag}(\lambda_1^{(n)},\cdots,\lambda_{D_y}^{(n)})$. Then with $\mathbf{G}=\sqrt{\mathbf{W}}\mathbf{P}^\top$, \eqref{multiuser_nec} becomes
$$
\sum_{n=1}^N\sqrt{\mathbf{W}}(\mathbf{W}+h_n^{-2})^{-1}h_n^{-2}\boldsymbol{\Lambda}^{(n)}(\mathbf{W}+h_n^{-2})^{-1}=\beta\sqrt{\mathbf{W}}
$$
Then for each diagonal element $w_d$ we have $w_d=0$ or
\begin{equation}\label{eq:weq}
\sum_{n=1}^{N}\frac{h_n^{-2}\lambda_d^{(n)}}{(w_d+h_n^{-2})^2}=\beta
\end{equation}
The LHS of the above equation is monotone decreasing with $w_d>0$, so we can easily find its solution by bi-section (or no solution when $w_d>0$). Denote the solution of \eqref{eq:weq} as $w_d(\beta)$ (assume $w_d(\beta)<0$ if no solution), then an local optimal value can be found with $w_d = \operatorname{max}(0,w_d(\beta))$. 
\end{IEEEproof}

Next, we tackle the more general case. The results in single-user case and the above theorem imply that we can set $\mathbf{P}$ as the eigenvectors of $\sum_{n=1}^Nm_n^2\mathbf{M}_n$, where $\{m_n\}_{n=1}^N$ can be seen as weights assigned to different users' tasks. Note that the different choices of $\{m_n\}_{n=1}^N$ won't affect the eigenvectors in Thm \ref{mutliuser_optimal}. And one of the most simple design of $\{m_n\}_{n=1}^N$ is to just set $m_n=1, n=1,\dots,N$. 

Here we also provide another design in high SNR case. When $\mathbf{G}\gg h_n^{-2}\mathbf{I}, \forall n$, the optimization goal can be approximated as follows:
\begin{equation}
\begin{aligned}
    & \sum_{n=1}^N \operatorname{Tr}(h_n^{-2}\mathbf{M}_n(\mathbf{G}^\top\mathbf{G}+h_n^{-2}\mathbf{I})^{-1}) \\
    \approx & \operatorname{Tr}((\sum_{n=1}^Nh_n^{-2}\mathbf{M}_n)(\mathbf{G}^\top\mathbf{G})^{-1})
\end{aligned}    
\end{equation}
The final expression can be seen as a single-user problem with $\mathbf{M}=\sum_{n=1}^Nh_n^{-2}\mathbf{M}_n$, which obtains optimal value via choosing basis as eigenvectors of $\sum_{n=1}^Nh_n^{-2}\mathbf{M}_n$. And this indicates that in high SNR case, we can set  $m_n=h_n^{-1}$. 

Similarly, in low SNR case, we can approximate the optimization goal as
\begin{equation}
\begin{aligned}
    & \sum_{n=1}^N \operatorname{Tr}(h_n^{-2}\mathbf{M}_n(\mathbf{G}^\top\mathbf{G}+h_n^{-2}\mathbf{I})^{-1}) \\
    \approx & \sum_{n=1}^N \operatorname{Tr}(h_n^{-2}\mathbf{M}_n(h_n^2\mathbf{I}-h_n^4\mathbf{G}^\top\mathbf{G})) \\
    = & \sum_{n=1}^N \operatorname{Tr}(\mathbf{M}_n)-\sum_{n=1}^N \operatorname{Tr}(h_n^{2}\mathbf{M}_n(\mathbf{G}^\top\mathbf{G}))
\end{aligned} 
\end{equation}
which can be optimized when $\mathbf{G}$ is the eigenvector with largest eigenvalues of $\sum_{n=1}^Nh_n^2\mathbf{M}_n$ --- and this also implicates that we can choose $m_n=h_n$.

To cover the two special cases considered above, we heuristically propose the following way to determine the weights $\{m_n\}$, i.e.,
\begin{equation}
    m_n=\frac{h_n}{h_n^{2}+\text{SNR}^{-1}_{\operatorname{min}}}
\end{equation}
where $\text{SNR}^{-1}_{\operatorname{min}}$ is defined as the minimum receiving SNR among all users. With normalized receiver noise power, it is 
$$
\begin{aligned}
    \text{SNR}_{\operatorname{min}} & =Eh_{\operatorname{min}}^2 \\
    h_{\operatorname{min}} & \triangleq \min(\vert h_1\vert,\cdots,\vert h_N\vert)
\end{aligned}
$$

Suppose we have obtained $\mathbf{P}$ through eigen decomposition of $\sum_{n=1}^Nm_n^2\mathbf{M}_n$. Substitute $\mathbf{G}=\sqrt{\mathbf{W}}\mathbf{P}^\top$ into \eqref{P2} and we get the following energy allocation problem with respect to $\mathbf{W}$:
\begin{equation}\label{prob:multiuser_energy}
    \min_{\{w_d\}} \ \sum_{d=1}^{D_y}(\sum_{n=1}^{N}\mathbf{p}_d^\top\mathbf{M}_n\mathbf{p}_d\frac{h_n^{-2}}{w_d+h_n^{-2}}) \quad \text{s.t.} \ \sum_{d=1}^{D_y}w_d\leq E
\end{equation}
We can solve the above problem by Lagrangian multiplier method, which can decompose it into $D_y$ sub-problems and each of them is convex to a $w_d$. 

\subsection{A Review of the Entire Structure}
Based on $\mathbf{G}=\sqrt{\mathbf{W}}\mathbf{P}^\top$, the decoder $\mathbf{F}_n$ is 
\begin{equation}
	\mathbf{F}_n = \boldsymbol{\Sigma}_{x_ny}\mathbf{P}h_n\sqrt{\mathbf{W}}(h_n^2\mathbf{W}+\mathbf{I})^{-1}
\end{equation}

There are two ways to understand the entire encoder-decoder structure. The first is on the level of raw data. To transmit $\mathbf{y}$, we allocate energy on different dimensions defined by $\mathbf{P}^\top$. At receiver side, the user first obtains an estimation $\hat{\mathbf{y}}$ of raw data with $\mathbf{P}h_n\sqrt{\mathbf{W}}(h_n^2\mathbf{W}+\mathbf{I})^{-1}$, and then completes its personal task as if $\mathbf{y}=\hat{\mathbf{y}}$. 

The second understanding is from the feature level. We can view each row of $\mathbf{P}^\top$ as a feature extractor, which extracts feature for further utilization of all tasks. The decoder actually first estimates the feature based on the received signal by $h_n\sqrt{\mathbf{W}}(h_n^2\mathbf{W}+\mathbf{I})^{-1}$. Afterwards, $\boldsymbol{\Sigma}_{x_ny}\mathbf{P}$ serves as a whole to directly reconstruct interested $\mathbf{x}_n$ from the estimated features. Both of the two understandings show that the wireless transceiver can be directly embedded into the original estimation process. 

\subsection{Geometric Illustration}
We provide a more explicit explanation on how the correlation among tasks are utilized in our method. The eigenvectors of $\sum_{n=1}^Nm_n^2\mathbf{M}_n$ are actually the right singular vectors of $(m_1\boldsymbol{\Sigma}_{x_1y}^\top,\cdots,m_N\boldsymbol{\Sigma}_{x_Ny}^\top)^\top$, since
\begin{equation*}
	\sum_{n=1}^Nm_n^2\mathbf{M}_n=(m_1\boldsymbol{\Sigma}_{x_1y}^\top,\cdots,m_N\boldsymbol{\Sigma}_{x_Ny}^\top)\times\begin{pmatrix}
		m_1\boldsymbol{\Sigma}_{x_1y} \\
		\vdots \\
		m_N\boldsymbol{\Sigma}_{x_Ny}
	\end{pmatrix}
\end{equation*}
So no matter how we choose $\{m_n\}$, the eigenvectors provide a set of basis for the row space of $
(\boldsymbol{\Sigma}_{x_1y}^\top,\cdots,\boldsymbol{\Sigma}_{x_Ny}^\top)^\top
$, which we denote as $\overline{\boldsymbol{\Sigma}_{xy}}$. 

As stated in single-user case, the row space of $\boldsymbol{\Sigma}_{x_ny}$ can represent the interest of user $n$. So the row space of $\overline{\boldsymbol{\Sigma}_{xy}}$ 
actually represents all users' interested information on $\mathbf{y}$. Besides, from the subspace view, when we say $m$-th task and $n$-th task are correlated, it means that the row spaces of $\boldsymbol{\Sigma}_{x_my}$ and $\boldsymbol{\Sigma}_{x_ny}$ are not orthogonal to each other (but their intersection subspace might be zero-dimensional). 

In single user case, the projection directions are determined by the orthogonal basis of the task-relevant subspace. And in multi-user case, to deal with the arbitrary correlation among the multiple tasks, we find basis for the entire space spanned by each user's $\boldsymbol{\Sigma}_{x_ny}$, even the basis may not belong to any user's subspace. 

To further reveal users' personalized interests on data, we can reconstruct each user's interested information (subspace) using these shared basis $\mathbf{p}_1,\cdots,\mathbf{p}_{D_y}$:
\begin{equation}\label{eq:multiuser_projection}
	\boldsymbol{\Sigma}_{x_ny}=\boldsymbol{\Sigma}_{x_ny}\mathbf{PP}^\top=\sum_{d=1}^{D_y}(\boldsymbol{\Sigma}_{x_ny}\mathbf{p}_d)\mathbf{p}_d^\top
\end{equation}

\begin{figure}
    \centering
    \includegraphics[width=\linewidth]{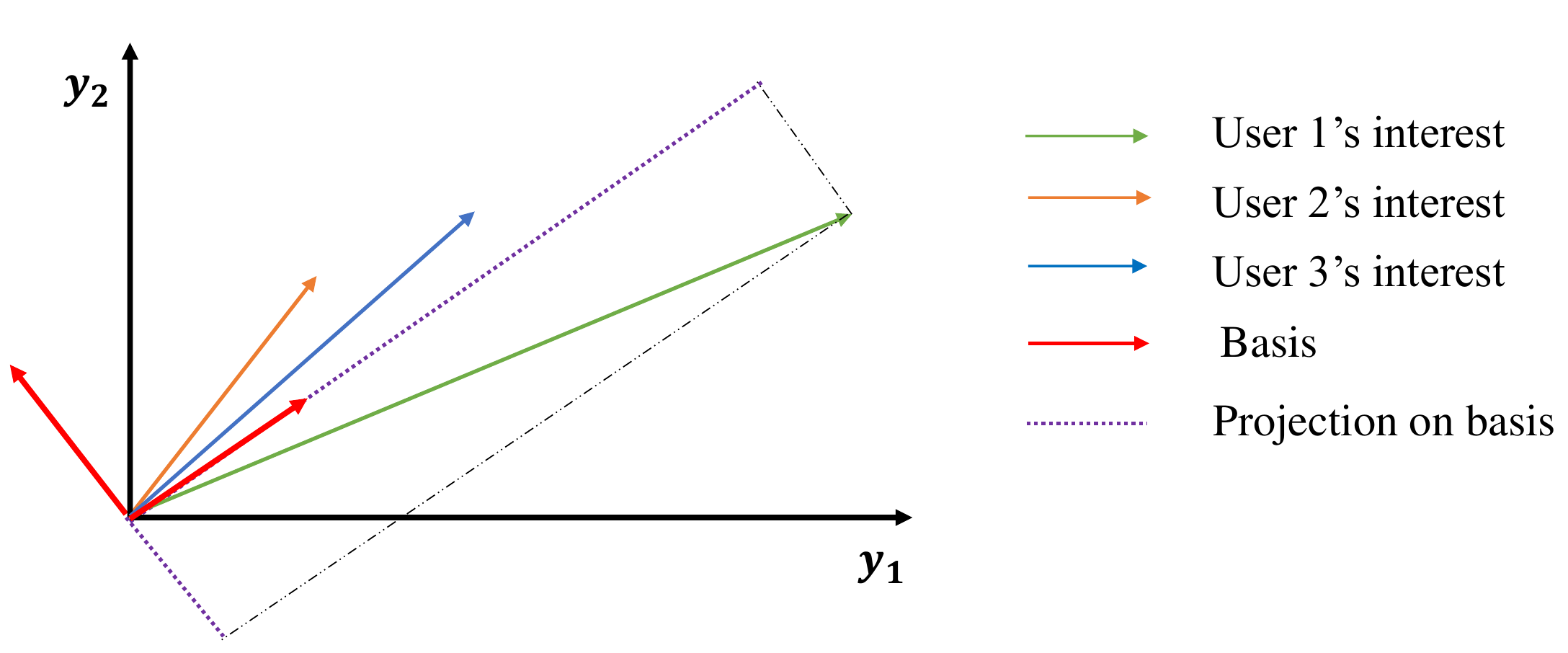}
    \caption{In multi-task case, we actually find the shared basis for all tasks, and represent each task using these basis. The projection on the basis shows how important such content is to the task.}
    \label{fig:my_label}
\end{figure}

The final optimization goal in \eqref{prob:multiuser_energy} is a weighted sum of the MSE of each feature. The weight, $\mathbf{p}_d\mathbf{M}_n\mathbf{p}_d^\top$, reveals the importance of the corresponding feature to the task --- the importance actually refers to how much average deviation in final estimation result may be brought by the estimation error of this feature. Interestingly, it can be verified that 
$$
\mathbf{p}_d\mathbf{M}_n\mathbf{p}_d^\top=\Vert\boldsymbol{\Sigma}_{x_ny}\mathbf{p}_d\Vert^2
$$
So the importance $\mathbf{p}_d\mathbf{M}\mathbf{p}_d^\top$ is reflected by the projection of a task's interested subspace on the basis. A simple illustration is given in Fig \ref{fig:my_label}. 

Finally, we highlight the row orthogonality of $\mathbf{P}^\top$, which makes the features extracted statistically independent. As a matter of fact, just to utilize the correlation of the tasks does not need the features to be independent. But such orthogonality makes the problem decomposable as \eqref{prob:multiuser_energy}, which greatly reduces the complexity of finding a good solution. 

However, the orthogonal basis can have many choices. And we only prove the optimality of basis found by SVD in special cases. For example, even we choose $\mathbf{P}=\mathbf{I}$, we can still allocate energy based on the importance of each dimension as problem \eqref{prob:multiuser_energy}. Other methods to find orthogonal basis such as Gram-Schmidt can also be applied here --- how the choice on basis will affect the result will be discussed in simulation part. 

\section{Application in Non-Linear Case}
In this section, we try to transfer conclusions learned in Gaussian linear case to non-linear tasks. We have shown that we can just embed the wireless transceiver into the original optimal estimator. So as the first step, we have obtain the original inference process. And we adopt the second explanation of transceiver design, that is, the estimator needs to extract the shared features from for all tasks and then each task is completed based on these features. What's more, these features should be independent and we require their importance to all tasks.

In general it is hard to obtain optimal non-linear estimators, so we constrain our eyes on estimators approximated by deep neural networks. Besides, to achieve the goal of extracting shared features, we adopt the hard parameter sharing in multi-task learning \cite{ruder2017overview} --- multiple tasks share a part of layers to obtain common features. And here we consider a relatively simple case --- we only consider neural networks consisting of linear fully connected layers and non-linear activation layers, i.e., multi-layer perceptron (MLP).

However, the independence of features is not a hard requirement in current neural networks, and it remains a challenging problem to ensure such independence in training a neural network. To achieve this, we propose a way to find independent features and weigh their importance after the neural network has been trained, which approximates the tasks by the first linear layers in task specific networks.

\subsection{Details in Encoder-Decoder Design}
Denote the shared network layers as $\Tilde{\mathbf{g}}(\cdot;\boldsymbol{\theta})$, and task specific layers as $\Tilde{\mathbf{f}}_n(\cdot;\boldsymbol{\theta}_n)$ for task $n$. And $\boldsymbol{\theta}$ and $\boldsymbol{\theta}_n$ are all parameters to be optimized based on:
\begin{equation}
\min_{\boldsymbol{\theta}, \{\boldsymbol{\theta}_n\}}\ \sum_{i=1}^M\sum_{n=1}^N\Vert\mathbf{x}_n^{(i)}-\Tilde{\mathbf{f}}_n(\Tilde{\mathbf{g}}(\mathbf{y}^{(i)};\boldsymbol{\theta});\boldsymbol{\theta}_n)\Vert^2
\end{equation}
where $\{(\mathbf{y}^{(i)},\{\mathbf{x}_n^{(i)}\}_{n=1}^N)\}_{i=1}^M$ is the training set. Then we denote the network layers with optimized parameters as $\Tilde{\mathbf{g}}(\cdot)$ and $\{\Tilde{\mathbf{f}}_n(\cdot)\}$. The total inference result can be written as
\begin{align}
    \mathbf{q} & = \Tilde{\mathbf{g}}(\mathbf{y}) \label{eq:sf}\\
    \hat{\mathbf{x}}_n & = \Tilde{\mathbf{f}}_n(\mathbf{q}) =  \overline{\mathbf{f}}_n(\mathbf{A}_n\mathbf{q}), \ n=1,\cdots,N
\end{align}
For the convenience of further derivation, we emphasize the linear input layer of the task specific networks and thus write $\Tilde{\mathbf{f}}_n(\mathbf{q})=\overline{\mathbf{f}}_n(\mathbf{A}_n\mathbf{q})$ \footnote{Without loss of generality, the first layer of task specific networks is linear layer.}. 

Then we focus on how to encode $\mathbf{q}$. Since the dimensions of $\mathbf{q}$ are not independent, we cannot direct allocate energy to each dimension based on importance. We simplify the problem by approximate the tasks by linear tasks, which also indicates the way to extract independent features. Denote $\hat{\mathbf{q}}$ as the estimated features after corrupted by communication noise, and to minimize final MSE we actually minimize the deviation in inference results caused by deviation in input features. Besides, for a neural network with linear layers and activation layers with bounded gradients, it is Lipschitz continuous. Thus we can assume  
$$
\Vert\overline{\mathbf{f}}_n(\mathbf{a})-\overline{\mathbf{f}}_n(\mathbf{b})\Vert \leq C_n\Vert\mathbf{a}-\mathbf{b}\Vert
$$ 
Then our optimization goal can be relaxed as
\begin{equation}
    \begin{aligned}
         \operatorname{E}[\Vert\overline{\mathbf{f}}_n(\mathbf{A}_n\mathbf{q})-\overline{\mathbf{f}}_n(\mathbf{A}_n\hat{\mathbf{q}})\Vert_2^2] 
         \leq  C_n\operatorname{E}[\Vert\mathbf{A}_n\mathbf{q}-\mathbf{A}_n\hat{\mathbf{q}}\Vert_2^2]
    \end{aligned}
\end{equation}
by which we have approximated original tasks by linear tasks of computing $\mathbf{A}_n\mathbf{q}$, so we can use results in linear case to design the  encoder of $\mathbf{q}$. Or equivalently, design encoder of $\Tilde{\mathbf{q}}=\boldsymbol{\Sigma}_{q}^{-1/2}(\mathbf{q}-\boldsymbol{\mu}_q)$ for tasks of computing $\mathbf{A}_n\boldsymbol{\Sigma}_{q}^{1/2}\Tilde{\mathbf{q}}$, where $\boldsymbol{\Sigma}_q$ and $\boldsymbol{\mu}_q$ are empirical covariance and mean value of the shared feature $\mathbf{q}$ computed based on the training set. 

We design the linear encoder as $\mathbf{G}=\sqrt{\mathbf{W}}\mathbf{P}^\top$. $\mathbf{P}^\top$ is chosen as right singular vectors of $(\mathbf{A}_1^\top,\cdots,\mathbf{A}^\top_N)^\top\boldsymbol{\Sigma}_q^{1/2}$.
Then the energy allocation matrix $\mathbf{W}=\operatorname{diag}(w_1,\cdots,w_{N_q})$ is obtained through optimizing the following problem:
\begin{equation}
\begin{aligned}
\min_{\{w_d\}} \ & \sum_{d=1}^{D_q}\sum_{n=1}^N\Vert \mathbf{A}_n\boldsymbol{\Sigma}_q^{1/2}\mathbf{p}_d\Vert^2_2\frac{h_n^{-2}}{w_d+h_n^{-2}} \\
\text{s.t.} \ & \sum_{d=1}^{D_q}w_d \leq E
\end{aligned}
\end{equation}



In the end, the whole procedure is listed as below:
\begin{enumerate}
    \item Compute shared features $\widetilde{\mathbf{q}}$
    \item Broadcast $\mathbf{s}=\mathbf{G}\widetilde{\mathbf{q}}=\operatorname{diag}(\sqrt{w_1},\cdots,\sqrt{w_{D_q}})\mathbf{P}^\top\widetilde{\mathbf{q}}$
    \item Estimate the features $\hat{\widetilde{\mathbf{q}}}_n=h_n\mathbf{G}^\top(\mathbf{G}\mathbf{G}^\top+\mathbf{I})^{-1}\mathbf{r}$, based on received signal $\mathbf{r}=h_n\mathbf{s}+\mathbf{u}_n$
    \item Compute $\hat{\mathbf{x}}_n=\mathbf{f}_n(\mathbf{A}_n\boldsymbol{\Sigma}_q^{1/2}\hat{\widetilde{\mathbf{q}}}_n+\mathbf{A}_n\boldsymbol{\mu}_q)$
\end{enumerate}

\begin{figure}[htbp]
    \centering
    \includegraphics[width=\linewidth]{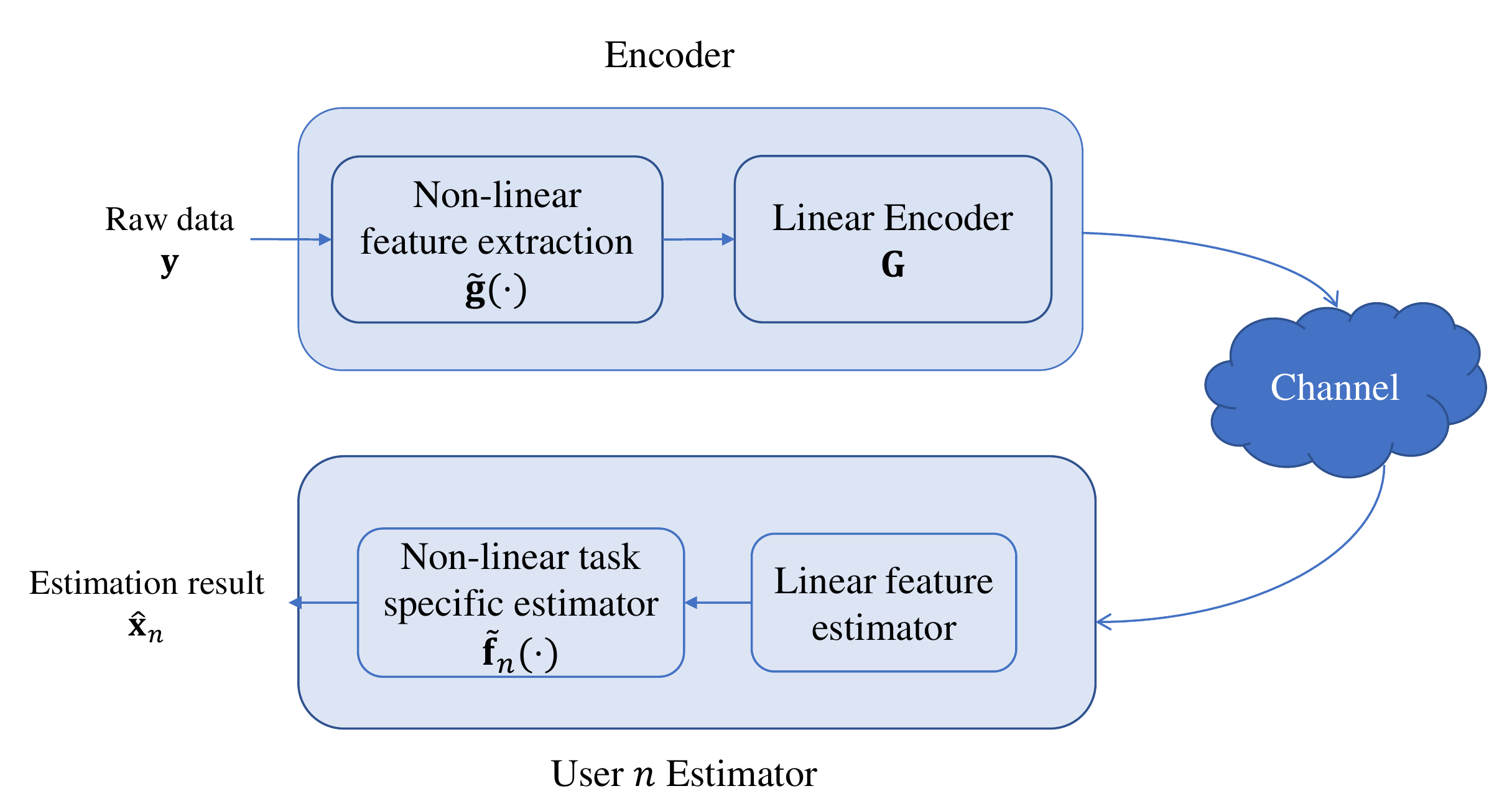}
    \caption{Total structure in non-linear problem.}
    \label{fig:structure}
\end{figure}

The total structure of encoder can be expressed as first extracting shared features using non-linear encoder and then allocating energy in a specific projection space of the features. And the decoder side first estimates the features using linear estimator and then feeds the result into subsequent non-linear estimators. We show it in Fig. \ref{fig:structure}.


\section{Numerical Results}
\subsection{Linear Estimation Task}
\subsubsection{Data Generation Model}
We test our method using synthetic data. We suppose $\mathbf{y}$ and $\{\mathbf{x}_n\}$ are generated following the model
\begin{equation}\label{eq:gen_data}
    \begin{aligned}
        \mathbf{y} & = \mathbf{C}\mathbf{z}+\mathbf{v} \\
        \mathbf{x}_n & = \mathbf{K}_n\mathbf{z}
    \end{aligned}
\end{equation}
where $\mathbf{z}\sim\mathcal{N}(\mathbf{0},\mathbf{I}), \mathbf{v}\sim\mathcal{N}(\mathbf{0},\mathbf{I}),\mathbf{z}\in\mathbb{R}^{D_z},\mathbf{C}\in\mathbb{R}^{D_y\times D_z},\mathbf{K}_n\in\mathbb{R}^{D_x\times D_z}$. And we generate $\mathbf{C},\{\mathbf{K}_n\}$ as Gaussian matrices. Besides, we can further constrain the rows of all $\mathbf{K}_n$ all lie in a low-dimensional subspace to facilitate the correlation among tasks, i.e., $\operatorname{Range}(\mathbf{K}_n^\top)\subset \mathcal{S}, \forall n$ and $\sum_{n=1}^N\operatorname{rank}(\mathbf{K}_n)<\operatorname{dim}(\mathcal{S})\leq D_z$. 

\subsubsection{Baseline Methods}
We compare our proposed joint precoding method with the optimal solution obtained by SDP and other two baselines: 
\begin{itemize}
    \item Time division multiplexing (TDM): Optimize encoder for each user separately and send user-specific signals one by one.
    \item Direct broadcast: Broadcast all dimensions of raw data with same power, i.e., set $\mathbf{G}=\eta\mathbf{I}$. 
\end{itemize}

\begin{figure}[ht]
    \centering
    \includegraphics[width=\linewidth]{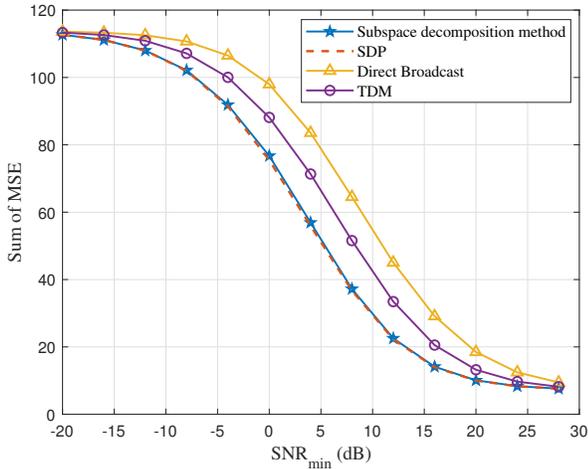}
    \caption{Results with linear estimation tasks.}
    \label{result:linear}
\end{figure}

\subsubsection{Improvements over baseline methods}
System parameters are set as $N=4,D_z=20,D_x=4,D_y=30,\operatorname{dim}(\mathcal{S})=8$. Users channel $\{h_n\}$ are generated from i.i.d Gaussian distribution. The results are shown in Fig. \ref{result:linear}. It can be seen that our method performs better than the other two heuristic baselines, and has similar performance compared with the optimal SDP based method. The complexity of our proposed method mainly lies with SVD, which is $\mathcal{O}(\operatorname{min}(N^2D_x^2D_y,ND_xD_y^2))$; while the complexity of the SDP method scales at $\mathcal{O}((ND_y^2)^3)$, which is much more expensive.

Directly broadcasting raw data ignores the fact that only a part of information is required by the tasks; while TDM scheme does not utilize the correlation among the interested contents of different tasks. Both of them transmit redundant information and thus are less energy efficient than our method.
\subsubsection{Effect of different orthogonal basis}
\begin{figure}[ht]
	\centering
	\includegraphics[width=\linewidth]{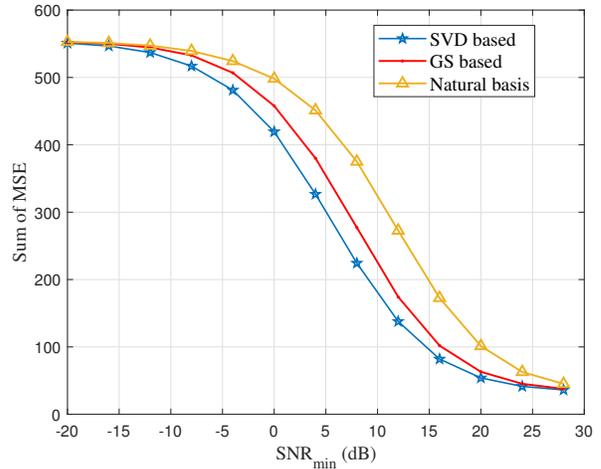}
	\caption{Comparison of different orthogonal basis.}
	\label{orthcomp}
\end{figure}

We also compare the effect of choosing basis from different methods. We choose orthogonal basis from SVD, Gram-Schmidt (GS) process and the natural basis (i.e., use identity matrix"). The result is shown in Fig. \ref{orthcomp}. The natural basis performs worst as it does not consider the requirement of the tasks, and the basis might not lie in the subspace defined by the tasks (row space of $\overline{\boldsymbol{\Sigma}_{xy}}$). The basis obtained from SVD and Schmidt process have similar performance but SVD is better in this simulation. How to determine the best set of basis is related with how to extract features in neural network based scenarios --- the best extractor in non-linear case is still worth further study. 

\subsection{Non-linear Estimation Tasks}
\subsubsection{Data generation and network architecture}
We simulate a system consisting of $3$ users, and we use the following model to generate raw data and users' interested parameters:
$$
\begin{aligned}
    \mathbf{x}_1 & = \begin{pmatrix} 2\sin(\mathbf{z}[0])+\operatorname{max}(\mathbf{z}[1],0)+0.5\exp(-\mathbf{z}[2]^2) \\
  \operatorname{max}(\mathbf{z}[1],0)+3\exp(-\mathbf{z}[2]^2) \\
   \sin(\mathbf{z}[0])-\exp(-\mathbf{z}[2]^2)\end{pmatrix}\\
   \mathbf{x}_2 & = \begin{pmatrix}
       0.3\mathbf{z}[4]^2+2\cos(\mathbf{z}[1]) \\
       -0.2\operatorname{max}(\mathbf{z}[1],0)+3\exp(-\mathbf{z}[2]^2)\\
       \sin(0.01\mathbf{z}[0])+\arctan(\mathbf{z}[3])
   \end{pmatrix}\\
   \mathbf{x}_3 & = \begin{pmatrix}
       2\mathbf{z}[1]+0.1\sin(\mathbf{z}[3]) \\
       2\sin(\mathbf{z}[0])-0.05\arctan(\mathbf{z}[3])+\vert\mathbf{z}[1]\vert \\
       \operatorname{max}(\mathbf{z}[1],0)
   \end{pmatrix} \\
    \mathbf{y} & = \mathbf{C}\mathbf{z}+\mathbf{v}
\end{aligned}
$$
where $\mathbf{z}\sim\mathcal{N}(\mathbf{0},\mathbf{I}), \mathbf{v}\sim\mathcal{N}(\mathbf{0},\mathbf{I}), \mathbf{C}\in\mathbb{R}^{20\times 8}$. 

The architecture of the network is listed in Table \ref{tab:my_label}. 
\begin{table}[ht]
    \caption{Network Architecture (From Bottom to Up)}
    \centering
    \begin{tabular}{c|c}
    \hline
        Shared Network &  Task Specific Network\\
        \hline
        \makecell[c]{$20\times 16$ Linear \\ Tanh \\$16\times 12$ Linear \\Tanh} & \makecell[c]{$12\times 8$ Linear \\Tanh\\$8\times 3$ Linear} \\
        \hline
    \end{tabular}
    \label{tab:my_label}
\end{table}
\subsubsection{Results}
\begin{figure}[htbp]
    \centering
    \includegraphics[width=\linewidth]{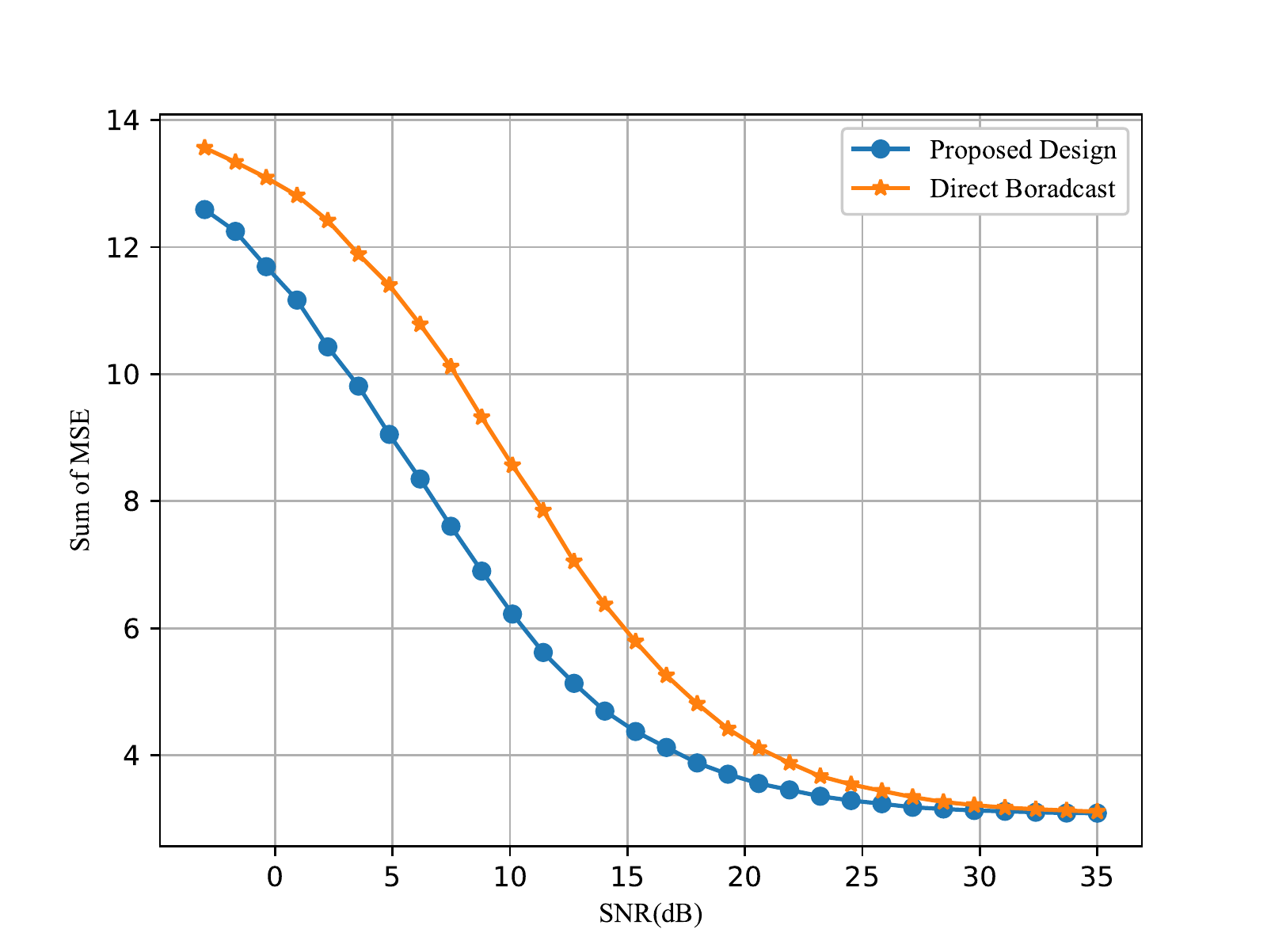}
    \caption{Results in non-linear case.}
    \label{fig:nl}
\end{figure}

We compare our method with directly broadcasting the whitened shared features (Fig. \ref{fig:nl}). The result is consistent with intuition --- since our method is able to allocate energy based on heterogeneous importance, it performs better. 

\section{Conclusion}
In this paper an analog precoding scheme is proposed for serving different estimation tasks required by multiple users. The core is to find the targeted information shared by multiple tasks and we achieve this from the view of subspace. Our method achieves almost the same performance compared with optimal solution. And the ideas behind our solution can be extended to more complex scenarios --- it proposes a two-stage method in designing multi-task oriented communication system: 
\begin{enumerate}
    \item extracting ``orthogonal" contents for all tasks jointly and weighting their importance to different tasks;
    \item allocating resource among these contents to maintain the transmission quality at different users based on importance.
\end{enumerate}
In linear case, our proposed method reached the first goal by SVD and in non-linear case, we attempt to extract shared feature with the help of neural network, but the orthogonality and importance ranking are still achieved by linear operation. How to complete the desired feature extraction in deep learning more effectively is worth further study. The second step is closely related to joint source channel coding (JSCC). JSCC under broadcast channel with heterogeneous service requirements is another direction of our future work. 








\bibliographystyle{ieeetr}
\bibliography{ref}

\end{document}